# Structural transformation and localization during simulated nanoindentation of a non-crystalline metal film

Yunfeng Shi and Michael L. Falk

Department of Materials Science and Engineering, University of Michigan, Ann Arbor, MI

48109-2136

Abstract

A simulation study demonstrates that localization can arise as the result of the breakdown of stable quasi-crystal-like atomic configurations. Samples produced at elevated quench rates and via more energetic processes contain a lower fraction of such configurations and exhibit significantly less pronounced localization and shorter spacing between bands. In the samples produced by the lowest quench rates localization is accompanied by the amorphization of material with initially quasi-crystal-like medium range order. This result is of particular significance in light of recent experimental evidence of local quasi-crystal order in the most stable of the bulk metallic glasses.

Non-crystalline materials do not naturally conform to unique low energy structures, and consequently their mechanical properties exhibit a high sensitivity to processing. Amorphous metals in particular can be produced by casting[1,2], quenching at elevated cooling rates[3] and extremely energetic non-equilibrium techniques such as ion beam processing. Micro- and nano-hardness result in shear band formation[4-10] accompanied by serrated mechanical response.

Plastic deformation arises from a qualitatively different physical process in non-crystalline metals than in poly-crystalline metals. Since dislocation glide is precluded by lack of a regular crystal lattice, activation of the rearrangement of a particularly oriented clusters of atoms, known as a shear transformation zones (STZ), mediates deformation[11,12]. These same mechanisms have been instrumental in devising constitutive theories that describe rheology near the glass transition[13] and understanding pressure or normal stress dependence in their mechanical response[14].

The potential causes of localization in these materials include softening due to STZ proliferation and elastic interactions between STZs[15-17]. Despite recent progress, there remains no method of characterizing the structural origin of the softening mechanism. As a result, while the most widely accepted theories[11,18,19] provide important physical insight, none of these can be used to accurately model the nucleation and propagation of shear bands or to characterize materials for use in structural applications, since the parameters in the current theories cannot unambiguously be attributed to specific atomic scale structures. Strain localization has been observed in molecular dynamics studies[20], but an atomic signature of the underlying structural transformation has not previously been identified.

To simulate the nano-indentation we have performed molecular dynamics simulation on a two-dimensional binary alloy. The simulation methods are as detailed in ref. 12. The alloy consists of two species, which we will refer to as S and L for small and large, interacting via a Lennard-Jones potential of the form

$$\phi_{ij} = 4\varepsilon \left[ \left( \frac{\sigma}{r_{ij}} \right)^{12} - \left( \frac{\sigma}{r_{ij}} \right)^{6} \right], \qquad (1)$$

where ε represents the bonding energy and $\sigma$ provides a length scale, the distance at which the interaction energy is zero. The SS and LL bond energies are half that of the SL bond energy, $\varepsilon_{SS} = \varepsilon_{LL} = \frac{1}{2}\varepsilon_{SL}$. The SS and LL length scales are related to the SL length scale by

$$\sigma_{SS} = 2\sigma_{SL}\ sin\left(\tfrac{\pi}{10}\right), \quad \sigma_{LL} = 2\sigma_{SL}\ sin\left(\tfrac{\pi}{5}\right). \quad (2)$$

We choose the reference length scale to be $\sigma_{SL}$ and the reference energy scale to be $\varepsilon_{SL}$. All the particles have the same mass, $m_0$. The reference time scale is $t_0 = \sigma_{SL}\sqrt{m_0/\varepsilon_{SL}}$. In order to make comparisons to experiments we will consider that for a typical material $t_0 \approx$ 1ps, and $\sigma_{SL} \approx$ 3Å.

This system was chosen because it exhibits a strong tendency toward amorphization while many other two-dimensional systems show a strong tendency to crystallize. It also exhibits both crystalline and quasi-crystalline ground states [21,22]. It has been proposed that such an underlying quasi-crystalline state stabilizes Zr based glasses[23,24]. In addition this system has been used to study quasi-crystal and amorphous thermodynamic and mechanical properties[22,25]. We chose our composition $N_L:N_S = (1+\sqrt{5}):4$ to be consistent with other studies of this system. $T_{MCT}$, the mode coupling temperature, was measured to be 0.325 $\varepsilon_{SL}/k$, where k is the Boltzmann factor. For the sake of comparison temperatures will be measured in units of this temperature which characterizes the onset of the glass transition.

The initial conditions were created by starting from supercooled liquids equilibrated above $T_{MCT}$. Subsequent to equilibration the temperature of the liquid was reduced to 9.2% of $T_{MCT}$. Sample I was cooled at a rate of $1.97 \times 10^{-6}$ $T_{MCT}/t_0$, corresponding to a quench over approximately 0.5 μs. Sample II was cooled at a rate of $0.98 \times 10^{-4}$ $T_{MCT}/t_0$, corresponding to a quench over approximately 10 ns. Sample III was quenched instantaneously by rescaling the particle velocities and then allowed to age for 100 $t_0$, approximately 0.1 ns. These samples were then tiled 5 across by 2 down to create a single slab of 200,000 atoms which formed the 285 $\sigma_{SL}$ or approximately 87 nm thick film. The indenter modeled by imposing a purely repulsive potential of the form $\phi_{Ii} = \varepsilon\left[(r_{Ii} - R_I)/(0.6\sigma_{SL})\right]^{-12}$, where $r_{Ii}$ is the distance between atom $i$ and the center of the spherical indenter of radius $R_I$. $R_I$ was chosen to be 250 $\sigma_{SL}$, approximately 75 nm. The indenter acts as a material of infinite stiffness with no surface friction. The slab was held between periodic boundaries on the right and left. The lower edge of the slab was held fixed as if the amorphous film were deposited upon a substrate of infinite stiffness. The indenter was lowered into the substrate under displacement control at a velocity of 0.001 $\sigma_{SL}/t_0$ to a displacement of 32.5 $\sigma_{SL}$, i.e. approximately 0.3 m/s to a depth of 9.75 nm. After reaching the maximum depth the indenter was held in position for 14000 $t_0$, approximately 14 ns, and unloaded at the same velocity.

Figure 1 shows the load displacement curves during loading and unloading for all three samples. The more gradually quenched the sample the higher the apparent hardness. The load displacement curve for sample III, the instantaneously quenched sample, is significantly smoother than the other two curves that each exhibit clear evidence of serrated flow. Serrations correspond to particularly abrupt events that may involve the formation of new shear band or slip along one or more existing shear bands. These simulations indicate that the degree of serration can depend sensitively on the processing of the glass.

In order to directly examine the onset of plastic flow below the indenter we have extracted the local strain in the vicinity of each atom in the glass. This was done using the procedure for extracting a best fit strain introduced in ref. 12. Fig. 2 shows images of the regions of high deviatoric strain in each sample at the maximum indentation depth. Sample I shows evidence of localized deformation underneath the indenter. The first shear bands nucleate at a displacement of 8 $\sigma_{SL}$, approximately 24 Å. This corresponded to an indentation depth of 4.9 $\sigma_{SL}$, approximately 15 Å. A few shear bands carry the majority of the plastic flow in sample I. Sample II exhibits the majority of deformation in a region immediately beneath the indenter, but the plastic flow is significantly more evenly distributed with smaller spacing between bands. Sample III exhibits deformation at a still finer scale with shear band spacing less than half that in sample I.

In order to extract information regarding the changes in structure that accompany deformation we have utilized the fact that the underlying quasi-crystalline ground state is composed of nine distinct atomic motifs consisting of an atom and its nearest neighbors[26]. We have analyzed the structure of the samples by determining if each atom resides in one of these motifs. Before indentation 74% of the atoms in sample I are in stable motifs; as compared to 58% in sample II, and 44% in sample III. During indentation the number of atoms in stable motifs decreases in samples I and II and increases in sample III. The spatial distributions of stable and unstable atoms after indentation are shown in Fig. 3. It is apparent that in sample I the shear bands coincide with regions in which atoms are no longer in stable motifs. In sample II a more subtle increase in unstable atoms is evident under the indenter. In sample III no correlated spatial inhomogeneities are evident under the indenter.

It has been proposed that small regions with quasi-crystalline order may be important for the stability of the amorphous state in some bulk metallic glasses[23,24]. In addition it has been asserted that deformation takes place primarily in the amorphous region of the material[27]. However recent experimental nanoindentation studies revealed possible signs of strain localization in some quasi-crystalline metals[28]. In the model system investigated we have observed that dispersed local quasi-crystal-like order plays a critical role in controlling the qualitative nature of deformation in the material. Gradually quenched samples appear to exhibit larger degrees of quasi-crystal-like order and increased strain localization. In the material that exhibits the strongest tendency to form shear bands deformation converts material from quasi-crystal-like to fully amorphous. This conversion of material may play a crucial role in the softening process that leads to the shear banding instability.

Acknowledgements

The authors acknowledge the support of the NSF under grant DMR-0135009 and the donors of the ACS Petroleum Research Fund for support under grant 37558-G.

References

1   A. Inoue, Acta Materialia **48** (1), 279 (2000).

2   W. L. Johnson, MRS Bulletin **24** (10), 42 (1999).

3   W. Klement, R. H. Willens, and P. Duwez, Nature **187** (4740), 869 (1960).

4   C. A. Schuh, A. S. Argon, T. G. Nieh, and J. Wadsworth, Philosophical Magazine **83** (22), 2585 (2003).

5   C. A. Schuh and T. G. Nieh, Acta Materialia **51** (1), 87 (2003).

6   R. Vaidyanathan, M. Dao, G. Ravichandran, and S. Suresh, Acta Materialia **49** (18), 3781 (2001).

7   W. J. Wright, R. Saha, and W. D. Nix, Materials Transactions **42** (4), 642 (2001).

8   Y. I. Golovin, V. I. Ivolgin, V. A. Khonik, K. Kitagawa, and A. I. Tyurin, Scripta Materialia **45** (8), 947 (2001).

9   T. Benameur, K. Hajlaoui, A. R. Yavari, A. Inoue, and B. Rezgui, Materials Transactions **43** (10), 2617 (2002).

10  W. H. Jiang and M. Atzmon, Journal of Materials Research **18** (4), 755 (2003).

11  A. S. Argon, Acta Metallurgica **27**, 47 (1979).

12  M. L. Falk and J. S. Langer, Physical Review E **57** (6), 7192 (1998).

13  M. L. Falk, J. S. Langer, and L. Pechenik, Physical Review E **70**, 011507 (2004).

14  C. A. Schuh and A. C. Lund, Nature Materials **2** (7), 449 (2003).

15  V. V. Bulatov and A. S. Argon, Modelling and Simulation in Materials Science and Engineering **2** (2), 167 (1994).

16  J. S. Langer, Physical Review E **64** (1), 011504 (2001).

17  L. O. Eastgate, J. S. Langer, and L. Pechenik, Physical Review Letters **90** (4), 045506 (2003).

18  F. Spaepen, Acta metall **25**, 407 (1977).

19  P. S. Steif, F. Spaepen, and J. W. Hutchinson, Acta Metallurgica **30** (2), 447 (1982).

20  F. Varnik, L. Bocquet, J.-L. Barrat, and L. Berthier, Physical Review Letters **90** (9), 095702 (2003).

21  F. Lancon, L. Billard, and P. Chaudhari, Europhysics Letters **2** (8), 625 (1986).

22  H. K. Lee, R. H. Swendsen, and M. Widom, Physical Review B **64** (22), 224201 (2001).

23  J. Saida, M. Kasai, E. Matsubara, and A. Inoue, Annales De Chimie-Science Des Materiaux **27** (5), 77 (2002).

24  K. Saksl, H. Franz, P. Jovari, K. Klementiev, E. Welter, A. Ehnes, J. Saida, A. Inoue, and J. Z. Jiang, Applied Physics Letters **83** (19), 3924 (2003).

25  M. L. Falk, Physical Review B **60** (10), 7062 (1999).

26  M. Widom, K. J. Strandburg, and R. H. Swendsen, Physical Review Letters **58** (7), 706 (1987).

27  J. Saida and A. Inoue, Scripta Materialia **50** (10), 1297 (2004).

28  V. Azhazha, S. Dub, G. Khadzhay, B. Merisov, S. Malykhin, and A. Pugachov, Philosophical Magazine **84** (10), 983 (2004).

[29] Avi animations of these simulation results are available via EPAPS.

**Figure Captions**

**Figure 1: The load displacement curves from three nanoindentation simulations performed on samples produced by quenching at a lower rate (I), higher rate (II) and instantaneously (III). The dashed line shows the elastic prediction for indentation in a material with the moduli of Sample I. Serrations are observed during indentation in Samples I and II but not III. For comparison $\sigma_{SL} \approx 3$Å and the indentation rate is approximately 0.3 m/s.**

**Figure 2: Visualization of the magnitude of the local deviatoric shear strain under the indenter at maximum depth in three samples produced at different quench rates[29]. Dark regions denote regions of high strain saturating at 40%. The scale bar in the upper right of the image is approximately 25 nm.**

**Figure 3: Visualization of the fraction of atoms with local quasi-crystal-like order. White denotes high levels of quasi-crystal-like order and black denotes absence of such order. The areas of low order in sample I clearly coincide with the locations of shear bands in Fig. 2. The scale bar in the upper right of the image is approximately 25 nm.**

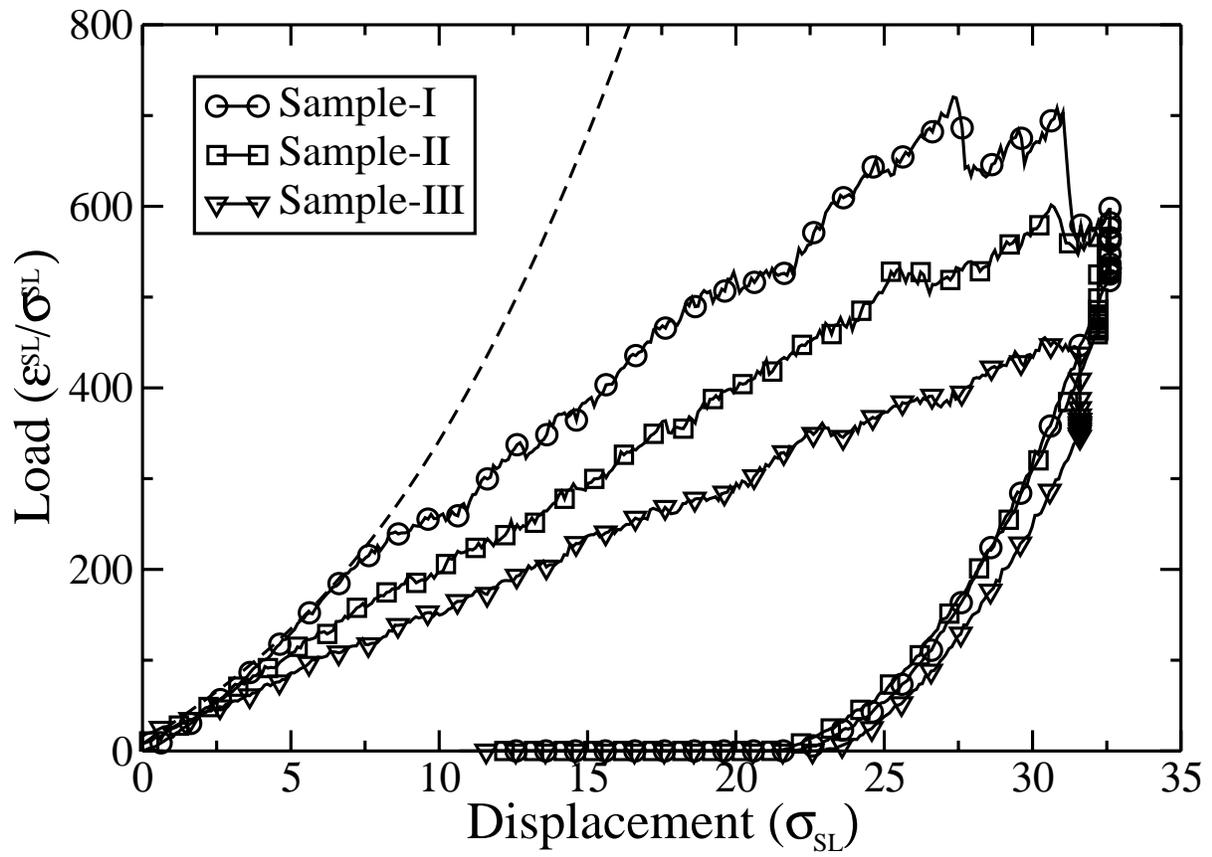

Figure 1

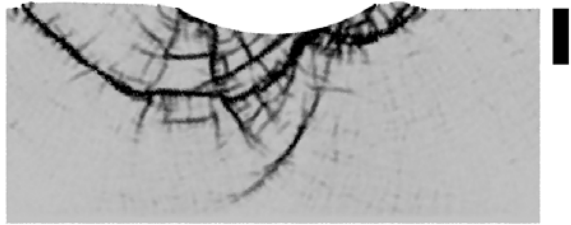
**Sample-I**

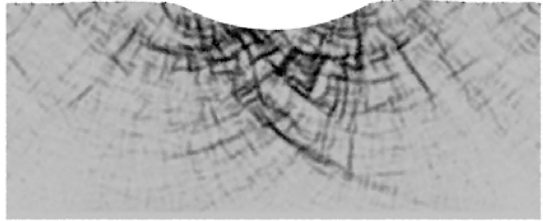
**Sample-II**

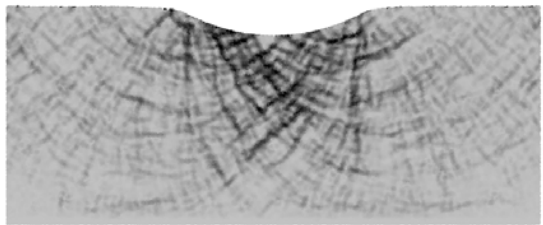
**Sample-III**

Figure 2

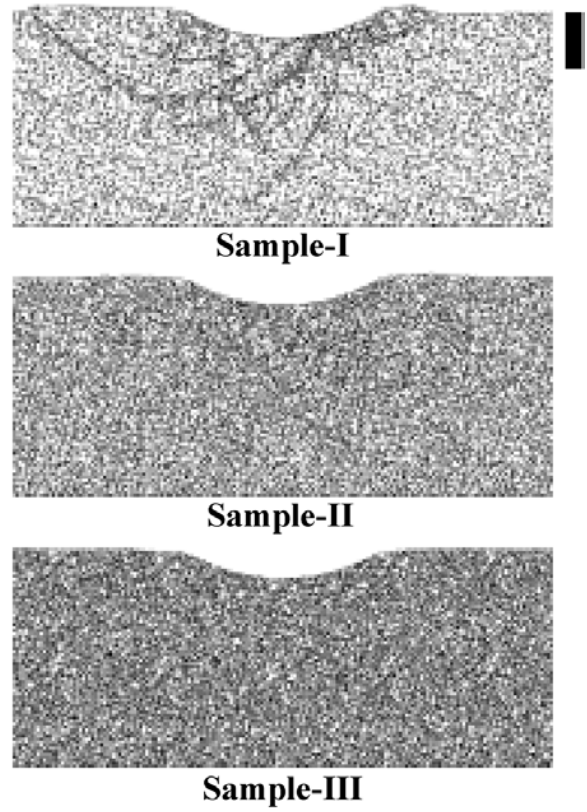

Figure 3